\documentclass{mn2e}
\usepackage{times}
\input{psfig.sty}

\newif\ifAMStwofonts

\title{The Connection Between Radio Quiet AGN and the High/Soft State
of X-Ray Binaries}

\author[Maccarone, Gallo \& Fender] {Thomas J. Maccarone, Elena Gallo
\& Rob Fender\\ Astronomical Institute ``Anton Pannekoek'', University
of Amsterdam, Kruislaan 403, 1098 SJ, Amsterdam, The Netherlands}

\date{}

\begin{document}

\maketitle

\label{firstpage}

\def\simlt{\mathrel{\rlap{\lower 3pt\hbox{$\sim$}}
        \raise 2.0pt\hbox{$<$}}}
\def\simgt{\mathrel{\rlap{\lower 3pt\hbox{$\sim$}}
        \raise 2.0pt\hbox{$>$}}}

\input epsf

\begin{abstract}

A large sample of AGN studied here shows a ``quenching'' of the radio
emission occurs when the luminosity is at a few percent to about 10\%
of the Eddington rate, just as is seen in the high/soft state of X-ray
binaries.  The result holds even when the sample of AGN includes no
Narrow Line Seyfert 1 galaxies (the systems most commonly suggested to
be the analog of the high/soft state).  This adds substantially to the
body of evidence that AGN show the same spectral state phenomenology
and related disc-jet coupling as the stellar mass accreting black
holes.  That the power law correlation between X-ray and radio
luminosity is the same in both AGN and X-ray binaries and extends
below $10^{-7} L_{EDD}$ strengthens the argument that there is no
fundamental difference between the low/hard state and the so-called
quiescent state in X-ray binaries.  We also discuss possible reasons
for the scatter in the radio to X-ray luminosity correlation in the
AGN.

\end{abstract}

\begin{keywords}
accretion,accretion discs -- galaxies:jets -- galaxies:Seyfert -- X-rays:binaries -- quasars:general
\end{keywords}

\section{Introduction}

Accreting black holes, whether stellar mass binary systems, or active
galactic nuclei, show many similarities.  Both emit radiation over
many decades in frequency.  Many, but not all, sources in each mass
range display evidence for disc accretion in the forms of thermal
emission and reflection off the thin disc.  Many, but not all, sources
in each mass range also have relativistic jets, typically seen through
radio observations (see Fender 2003 for a review of the radio
properties of X-ray binaries and chapter 2 of Krolik 1999 for a brief
introduction to the broadband properties of AGN).  Quenching of these
radio jets - defined to be a rather sudden drop in the radio emission
- has been known about for some time, both in certain classes of AGN
(Epstein et al. 1982) and in certain X-ray binaries (Tananbaum et
al. 1972; Waltman et al. 1996; Harmon et al. 1997; Fender et
al. 1999).

X-ray binaries thought to contain black holes have at least three
spectral states. The {\it low/hard state} shows X-ray emission well
fit by either a power law with an exponential cutoff at about 200 keV
or a thermal Comptonisation model with a temperature of about 70 keV,
along with a weak or undetected thermal component associated with a
geometrically thin accretion disc (Zdziarski 2000 and references
within).  A so-called ``quiescent state'' has been suggested to exist
in the black hole candidates, but in reality, the X-ray and radio
properties of low luminosity black hole candidates form a continuum
down to the lowest observable luminosities (Gallo, Fender \& Pooley
2003 - GFP03).  The spectrum of the {\it high/soft state} is dominated
by thermal component thought to arise in a geometrically thin,
optically thick accretion disc (Shakura \& Sunyaev 1973; Novikov \&
Thorne 1973) and also exhibits a weak power law tail without an
observable cutoff (e.g. Gierli\'nski et al. 1999).  The {\it very high
state} displays the same two components, but with the steep power law,
rather than the thermal component dominating the total flux (see
e.g. Miyamoto et al. 1991).  Steady radio emission, which can be
proved by brightness temperature arguments to come from a region
larger than the binary separation of the system, and hence likely from
a jet, has been found in the low/hard state, and strong radio flares
have been seen during the very high/flaring state (Fender 2003).
Neither strong nor steady radio emission has ever been detected in the
high/soft state.  For a review of spectral states of X-ray binary
black holes, see Nowak (1995) and references within.

Models for the hard non-thermal emission generally invoke
Comptonisation in a low optical depth (i.e. $\tau\simlt$1 in the
low/hard state and $\tau\simlt$5 in the very high state) geometrically
thick medium (e.g. Thorne \& Price 1975; Shapiro, Lightman \& Eardley
1976), although it has also been suggested that the low/hard state's
X-ray emission may be optically thin synchrotron emission from a jet
(Markoff, Falcke \& Fender 2001- MFF01).  Nonetheless, there is
general agreement that the high/soft state, the one state thought to
be geometrically thin is the one state where the radio emission is
suppressed substantially.  Theoretical studies suggest that jet
production should be suppressed in thin discs due to the lack of
poloidal magnetic fields (Livio, Ogilvie \& Pringle 1999; Khanna 1999;
Meier, Koide \& Uchida 2001).

Radio emission from active galactic nuclei is also correlated with the
X-ray spectral properties of the system, and with the geometry of the
inner accretion flow.  Radio loud AGN typically have harder X-ray
spectra than radio quiet AGN (Elvis et al. 1994; Zdziarski et
al. 1995).  Double-peaked optical emission lines are more often
detected in the radio loud AGN with broad lines than in their radio
quiet broad line counterparts (e.g. Eracleous \& Halpern 1994); it has
been suggested that these lines are reprocessed flux from a large
scale height X-ray emitting region (e.g. Chen, Halpern \& Fillipenko
1989).

More detailed evidence that AGN may follow the same spectral state
behavior as the black hole binaries has been scarcer.  Analogies have
been drawn (1) between the Narrow Line Seyfert 1 galaxies and the
high/soft state (Pounds, Done \& Osborne 1995; McHardy et al. 2003),
(2) among low luminosity AGN, Fanaroff-Riley (FR) I galaxies
(i.e. core dominated radio galaxies - see Fanaroff \& Riley 1974) and
the low/hard state (Meier 2001; Falcke, K\"ording \& Markoff 2003 -
FKM03) and (3) between FR II galaxies (i.e. lobe dominated radio
galaxies) and the very high state (Meier 2001).  The similarities
between the jet ejection events in 3C120 and GRS 1915+105 (Marscher et
al. 2002) can be interpreted as evidence for an analogy between the FR
II galaxies and the very high state.  Finally, GFP03 found that the high
luminosity ``transient'' jets are likely to have higher velocities
than the low luminosity ``steady'' jets.  This is a theoretical
prediction of at least one model for explaining the FR I/II dichotomy
(Meier 1999), and seems to be supported by the fact that FR I jets
tend to be double sided, while FR II jets tend to be single sided
(e.g. Hardcastle 1995), providing additional evidence in favor of
connections between FR I jets and the low/hard state, and between FR
II jets and the very high state.

One thing that has not been clear is whether the high/soft state might
exist for a broad range of masses of active galactic nuclei.  The
Narrow Line Seyfert 1 galaxies (NLSy1's) which represent the most
convincing high/soft state analogs have generally been found to
contain black holes at the low mass end of the AGN mass spectrum
(i.e. typically less than about 3$\times10^6 M_\odot$) and to be
accreting at tens of percent of the Eddington limit.  It has been
suggested that the radio quiet quasars may be higher mass examples of
the high/soft state (Merloni, Heinz \& Di Matteo 2003 - MHD03).

If one accepts a few reasonable, albeit unproven, assumptions about
the luminosities and mechanisms for state transitions in accreting
black hole systems, one can show that a high/soft state might not
exist above the mass limits found in the NLSy1's; Meier (2001)
suggests that the state transitions for the low/hard state to the
high/soft state are caused by transitions from an advection dominated
accretion flow (ADAF - see e.g. Ichimaru 1977; Esin, McClintock \&
Narayan 1997) to a standard thin disc, while the transitions to the
very high state occur when the thin disc becomes radiation pressure
dominated.  Since the ADAF to thin disc transition should occur at a
fraction of the Eddington luminosity independent of mass, while the
luminosity in Eddington units where the thin disc becomes radiation
pressure dominated goes as $M^{-1/8}$ (Shakura \& Sunyaev 1976), one
might expect the thin disc state to disappear when radiation pressure
domination sets in below the luminosity for the transition from ADAF
to thin disc; a similar, but less detailed argument for the same
effect had previously been laid out in Rozanska \& Czerny (2000).  The
soft-to-hard state transitions for black holes in binary systems are
generally found to occur at about 2\% of the Eddington limit
(Maccarone 2003), although the spectral states do show a hysteresis
effect (Miyamoto et al. 1995; Smith, Heindl \& Swank 2002; Maccarone
\& Coppi 2003), and the transition from the hard state to the soft
state can sometimes occur at luminosities about 4 times as high.  The
very high state seems to set in at about 20-30\% of the Eddington
luminosity, and from inspection of the classification table in Miller
et al. (2001), one can see that this state can also show hysteresis
effects in its transition luminosities and can exist down to about
10\% of the Eddington luminosity.  Given a factor of five ratio for a
10 $M_\odot$ black hole between where the very high state evolves into
the high/soft state and where the high/soft state evolves into the
low/hard state, one might then expect that no high/soft state systems
should exist for black holes more massive than about $4\times10^6$
solar masses.

Other models for the state transitions invoking whether the bulk of
the power in the accretion flow is dissipated as thermal energy or as
magnetic reconnection events can have the same scaling with mass for
both the soft/hard and the soft/very high state transitions (Merloni
2003).  Searching for the high/soft state analog in higher mass AGN
systems is thus a rather critical step in producing a unified
understanding of accretion processes in black hole systems.  In this
Letter, we will show that AGN accreting in the range of 2-10\% of the
Eddington luminosity show a ``quenching'' in their radio emission
similar to that found in Cygnus X-1 and GX 339-4 in their high/soft
states (Tananbaum et al. 1972; Fender et al. 1999; GFP), and that more
generally, the presence of a hard X-ray spectral component and radio
emission are well correlated (e.g. results from GRS 1915+105 in Harmon
et al. 1997; Klein-Wolt et al. 2002).  This similarities supports the
picture where (1) the high/soft state exists for AGN of a rather wide
range of masses and (2) that this high/soft state occurs at the same
range of X-ray luminosities as for the Galactic stellar mass black
hole candidates.

\section{Data and analysis}
The correlation between radio luminosity and broadband X-ray
luminosity found in GFP03 has been generalized for black holes of all
masses to be a $L_{R}-L_{X}-M$ correlation, through a
multi-dimensional analysis by MHD03 and through the application of a
theoretically predicted mass correction by FKM03.  Considerable
(i.e. several orders of magnitude) scatter does remain in the AGN
sample when this correlation is applied, and the difference cannot be
wholly due to measurement errors - additional parameters such as the
black hole spin might have a major effect on the radio luminosity.

The exact relation found by MHD03, which we have re-expressed in
Eddington units, is:
\begin{equation}
{\rm log} \frac{L_R}{L_{EDD}} = 0.60 {\rm log} \frac{L_X}{{L_{EDD}}} + 0.38 {\rm log}\frac{M}{M_\odot} - 7.33.
\end{equation}

We take the data used here from the compilation of MHD03.  The sample
includes AGN for which there are good mass, X-ray luminosity, and
radio luminosity measurements.  We exclude the sources in their sample
for which there is an upper limit rather than a measurement of one of
the three important quantities.  We then correct the radio luminosity
for the mass term as in equation (1), and plot the corrected radio
luminosity in Eddington units versus the broadband X-ray luminosity in
Eddington units in Figure 1 (a figure similar to Figure 7 of MHD03 and
Figure 3b of FKM03).

The X-ray luminosities have been multiplied by a factor of 4.8 as an
estimated ``broadband'' correction, assuming a $\Gamma=1.8$ power law
spectrum extending from 10 eV to 100 keV, as compared with the 2-10
keV range over which the luminosities have been tabulated.  This is
not quite a ``bolometric'' correction - our goal is to make a
comparison with the X-ray binaries for which RXTE's broadband
spectroscopy allows us to observe most of the X-ray emission.  We thus
do not wish to include the contribution from a component at
wavelengths longward of the accretion disc's peak, such as the radio
jet or a far infrared bump which may be partly due to AGN induced star
formation or slow reprocessing of AGN photons and may reflect the
past, rather than the present luminosity.  On the other hand, we do
wish to correct for the fact that the thin accretion discs of bright
AGN emit primarily in the optical and UV bands, and not in the X-rays.
From this point on, when we refer to the ``broadband luminsity'', we
mean the emission from the disc-corona system, as estimated by the
2-10 keV X-ray luminosity with a correction factor.

Low luminosity AGNs (i.e. those below about 1\% of the Eddington
luminosity) tend not have the ``big blue bump'' associated with the
emission from a thin disc, and this broadband correction factor is
well within the range generally accepted (e.g. Ho 1999), although our
correction factor is a bit smaller, because we correct only for the
high energy part of the spectrum associated with the disc-corona part
of the accretion flow.  For the brighter AGN, where the disc emission
is stronger (in agreement with the analogy to X-ray binaries), this
broadband correction may be a bit too low, and instead a broadband
correction of a factor of about 15-20 may be more reasonable.

\begin{figure*}
\vskip -1.5cm
\epsfxsize=7 cm \epsfysize=8.6 cm
\epsfbox{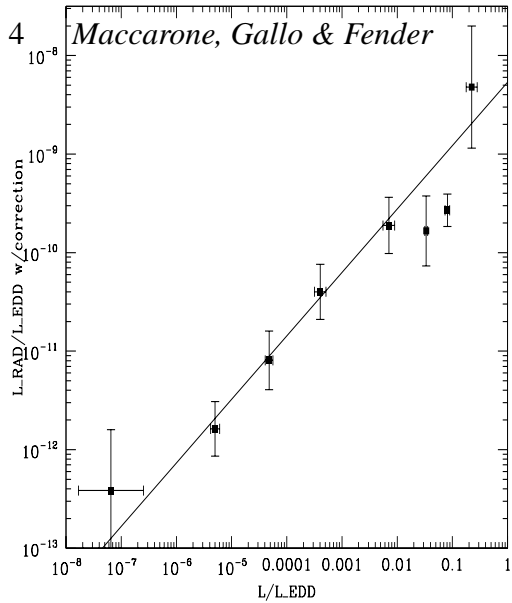}\epsfxsize=7 cm \epsfysize=8.6 cm
\epsfbox{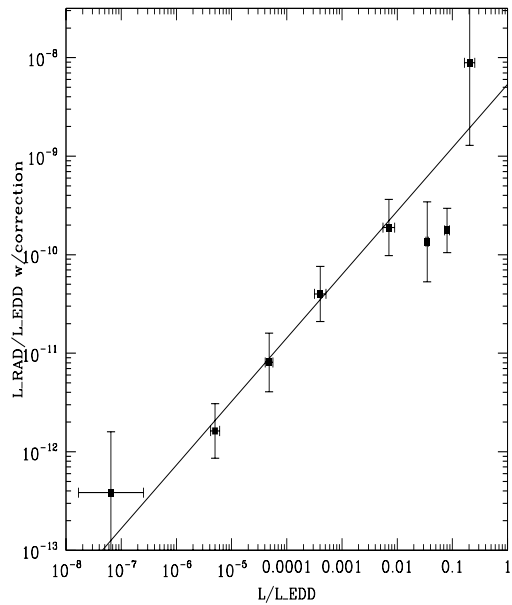}
\caption{Left:The binned corrected AGN, including the narrow line
Seyfert 1s.  Right: The same as left, but with the NLSy1s removed from
the sample.The line shows the best fit to the data (with the bins
thought to represent the high/soft state excluded from the fit): ${\rm
log} L_{R,corr}/L_{EDD} = (0.64 \pm 0.09) {\rm log} L_X/L_{EDD} -
(8.26 \pm 0.40)$.  The different normalization from the MHD03 results
comes mostly from the broadband correction.}
\end{figure*}

We have also re-binned the available data for the X-ray binaries where
simultaneous radio and X-ray points exist and where there are good
mass measurements for the systems.  The data set is described in
GFP03, and references within; we have included data for the following
low/hard state sources: GRO J0422+32, XTE J1118+480, 4U 1543-47, XTE
J1550-564, Cyg X-1, V404 Cyg \& GX 339-4.  We have included data from
GRS 1915+105 and the transient source sample of Fender \& Kuulkers
(2001) for the ``very high state.''  The transient source points are
not based on strictly simultaneous data, and hence may have some
systematic errors introduced, but ignoring these data points does not
change the results substantially, because the points lie very close to
the data for GRS 1915+105 and because most of the points in the high
luminosity bin come from GRS 1915+105 in any case.  As in the AGN
case, we have excluded points where the data are upper limits (for
fluxes) or lower limits (for masses), and we have applied the mass
correction from MHD03 to the data.  We have assumed a mass of 6
$M_\odot$ for GX 339-4, the mass function measured by Hynes et
al. (2003) and a distance of 4 kpc (Zdziarski et al. 1998), but we
note that this is a lower limit on the mass and not an actual
measurement.  We must include this source despite its not having an
actual mass measurement because it is the only X-ray binary with
simultaneous radio and X-ray data at the lowest luminosities.  We have
also tested the correlation assuming a mass of 9 $M_\odot$ and have
found that the results are not changed substantially.

In Figure 2, we have over-plotted the binned binary data with the data
from the AGN.  We have used the same binning ranges for the X-ray
binary sample as for the AGN sample, but we note that there are no
simultaneous radio and X-ray observations of X-ray binaries below
about $10^{-6} L_{EDD}$ and very few in the range around $10^{-3}
L_{EDD}$.  Also, there are relatively few points in our X-ray binary
sample very close to 10\% of the Eddington luminosity because the
radio data consists primarily of upper limits at this luminosity.  We
have also re-fit the AGN data without including the two ``quenched''
bins in the correlation and we find that:
\begin{equation}
{\rm log} \frac{L_{R,corr}}{L_{EDD}} = (0.64 \pm 0.09) {\rm log} \frac{L_X}{L_{EDD}} - (8.26 \pm 0.40),
\end{equation}
values that correspond much more closely with the X-ray binary
correlation found in GFP03.  We note here that the low luminosity AGN
and X-ray binaries show a very similar trend, and that the AGN
correlation extends several orders of magnitude lower in luminosity
than does the X-ray binary correlation.  By analogy, this bolsters
arguments that suggest that the quiescent state of X-ray binaries is
merely an extension of the low/hard state, and that the jet will begin
to dominate the total accretion power at very low luminosities (see
e.g. Fender, Gallo \& Jonker 2003).  We also show in Figure 3 the
ratio between the data points and the best fit to the non-quenched
data.  The AGN points with broadband luminosity at 3\% \& 10\% of the
Eddington luminosity are factors of $\sim5$ (i.e. 1.7$\sigma$ and
3.5$\sigma$, respectively) below the correlation.  We note that due to
the application of a broadband correction factor more appropriate to
lower luminosity AGN, the 3\% and 10\% of Eddington values are likely
to be underestimates by a factor of a few, and the points above a few
percent of the Eddington limit in the plots should be moved a bit to
the right.  Sliding the points to the right would push them a bit more
below the curve, so the quenching may actually be a bit stronger and
more statistically significant the the factor of $\sim5$ estimate.

\begin{figure}
\epsfxsize=2.8 in \epsfbox{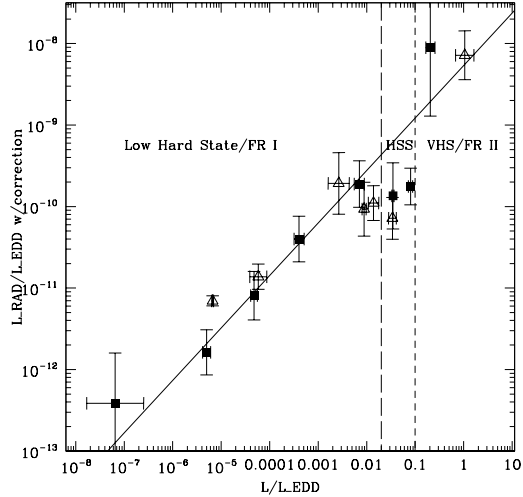} 
\caption{The same as Fig. 1b, with the X-ray binaries included.  The
open triangles represent the X-ray binaries.  The long-dashed vertical
line indicates the transition luminosity between the high/soft state (HSS)
and the low/hard state as measured in Maccarone (2003) and also is
very close to the transition luminosity between FR I \& II galaxies as
determined by Ghisellini \& Celotti (2001).  The short-dashed vertical
line indicates the estimated state transition luminosity between the
high/soft state and the very high state (VHS).  The fit to the data is the
same as that presented in Figure 1.}
\end{figure} 

\begin{figure}
\epsfxsize=2.8 in \epsfbox{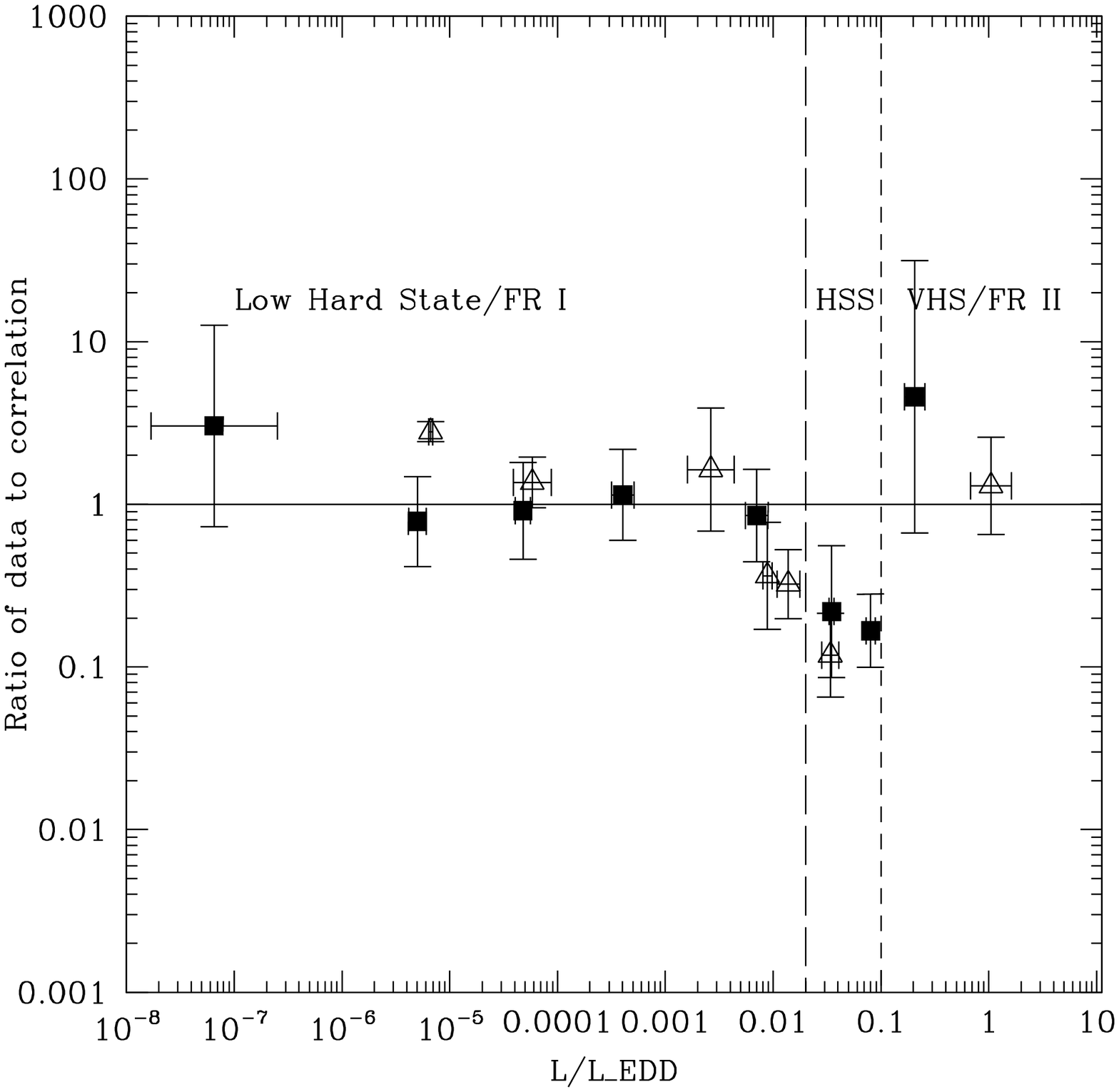}
\caption{The ratio of the data in Figure 2 to the best fit to the data.}
\end{figure}

\section{Discussion} 
The high/soft state appears to exist in AGN of all masses in the
sample which have broadband luminosities of order 5-10\% $L_{EDD}$.
The effect is not sensitive to whether the systems classified as
NLSy1's are included in the sample.  The masses of the mean AGN in the
bins where the downturn is seen are $5\times10^7 M_\odot$ and
$6\times10^7 M_\odot$, clearly above the typical masses for the
NLSy1's.

The observed correlations also help to underscore the importance of
considering the radio-to-X-ray luminosity ratio as a measure of the
radio loudness in addition to the more traditional radio-to-optical
luminosity ratio.  Such an approach does complicate comparisons with
some of the historical literature on AGN and optical surveys of AGN
tend do to be wider and deeper than those in the X-rays. Still the
optical emission in AGN sometimes is thermal emission from the
accretion disc and sometimes is non-thermal emission from a jet; the
relative contributions have dependences on black hole mass, redshift,
viewing angle and Eddington fraction.  The hard X-rays should always
be dominated by emission from the Compton corona (unless one adopts an
emission model such as that of MFF01 or Harris \& Krawczynski 2002, in
which case the hard X-ray emission should often be dominated by the
jet).  The greater homogeneity of emission mechanisms in the X-rays
contrasted with the optical makes correlations discovered in the
X-rays easier to interpret.  Furthermore, correlations made in the
X-ray will be easier to compare with correlations found in the stellar
mass black hole candidates.

In hindsight, it is not surprising that the high/soft state exists at
roughly the same luminosity for AGN as it does for X-ray binaries.
The well-known work of Ledlow \& Owen (1996) showed that the FR I
radio galaxies lie systematically below the FR II radio galaxies in a
plot of radio power versus $R$ magnitude.  Using empirical scaling
relations between the radio power and bolometric luminosity, and
between the $R$ magnitudes and black hole masses, Ghisellini \&
Celotti (2001) found that the dividing line corresponds to about 2\%
of the Eddington luminosity.  That the scaling relations used by
Ghisellini \& Celotti (2001) have considerable scatter, while there
are very few sources on the ``wrong'' side of the dividing line
between FR I and II galaxies seems may be taken as evidence that there
is intrinsically a gap where there are no strong radio galaxies that
would have been included in the 3C sample used in the Ledlow \& Owen
(1996) diagram.  The gap, where the high/soft state AGN exist in
reality, is then filled in by the low/hard state/FR I systems
scattered upwards and very high state/FR II systems scattered
downwards.  Understanding where is the error in the theoretical
predictions (Rozanska \& Czerny 2000;Meier 2001) that soft states
would not exist for such mass mass black holes remains an open
question; indeed many other mechanisms for producing state transitions
apart from those discussed above also predict a roughly $M^{-1/8}$
dependence of the state transition luminosities (e.g. Merloni 2003).

The quenching of the radio jets in the high/soft state range of
luminosities seems not to be as extreme for the AGN as it is in the
stellar mass systems; the AGN show a drop of a factor of only about 10
in the high/soft state. In the stellar mass black holes, the radio
luminosity drops by a factor of at least 30-50 from the low/hard state
correlation's extrapolation (Fender et al. 1999; Corbel et al. 2001).
Probably this is partly from contamination of the high/soft state AGN
luminosity range due to measurement errors on the masses of the black
holes in the AGN and possibly also the hysteresis effects seen in the
binary systems.  Also, our initial broadband correction underestimates
the real spectral correction factors for sources in this range,
leading to an additional underestimate of the quenching effect.

The $L_X-L_R$ correlation in MHD03 exhibits about 3 orders of
magnitude of scatter.  At least one order of magnitude is likely to
come from the use of the velocity dispersion-black hole mass technique
to measure most of the masses (e.g. Merritt \& Ferrarese 2001), but
this is unlikely to explain everything.  An excellent candidate for
the additional scatter would be black hole spin effects, since the
black hole spin may affect jet power either directly, if the jet is
the result of the extraction of black hole rotational energy
(Blandford \& Znajek 1977) or indirectly, if the jet is powered by the
rotational energy of the inner disc, which should be larger for a
rotating black hole (Blandford \& Payne 1982).  Given that the high
mass stars which are the progenitors of stellar mass black holes have
angular momenta much larger than the maximum angular momenta for black
holes of the same mass, it would not seem too unreasonable for all
stellar mass black holes to be rapidly rotating, as is suggested by
some models for the high frequency quasi-periodic oscillations in the
black hole binaries (Rezzolla et al. 2003).  On the other hand, black
hole-black hole mergers may contribute substantially to the spin
evolution of the black holes in AGN and would tend to reduce the spins
of most of the black holes produced in the mergers (Merritt 2002;
Hughes \& Blandford 2003).  It would hence not be too surprising if
the black holes in AGN show a much larger range of spins and of jet
power at a given mass and luminosity.  Testing this hypothesis may be
possible through iron line spectroscopy with the planned
Constellation-X mission, as iron lines have proved to be a powerful
diagnostic of spin in AGN (see e.g. Wilms et al. 2001).  It was found
in MHD03 that the scatter in the correlation is reduced by eliminating
sources in this range; we also find that the eliminating them makes
slope of the correlation in the AGN closer to that found in the X-ray
binaries.

More broadband spectroscopy should be undertaken on the putative
high/soft state AGN to determine if the systems are truly identical
to their lower mass counterparts; the work on NGC 4051 (McHardy et
al. 2003) shows that there may be systematic spectral differences
between the otherwise rather similar systems, as this system shows
similar variability characteristics to the high/soft state of X-ray
binaries, and a strong soft quasi-thermal component, but shows a
substantially harder power law tail.  This should be possible for the
brightest sources with a combination of observations from INTEGRAL and
from ground based optical telescopes.  

A start on this investigation can be made with the existing data.
Numerous studies of large samples of AGN suggest that the typical
spectral index $\alpha\simeq$-0.75, where $\alpha$ is defined by
$F(\nu)\propto$$\nu^\alpha$ (e.g. Wilkes \& Elvis 1987; Nandra \&
Pounds 1994; Lawson \& Turner 1997 - LT). From examining the
individual spectra of the sources in the 2-10\% of $L_{EDD}$ range, we
find that there are 15 sources, 3 of which are NLSy1's and hence are
known to have strong soft X-ray excesses.  Of the 12 non-NLSy1's, one
is a Seyfert 2 galaxy in which the hypothesis of a strong soft X-ray
excess is very difficult to test because of absorption, and one of the
quasars, PG 0844+349 also shows evidence of rather strong absorption.
One system (PG 1229+204) was rather faint and no spectral fit is
available in the literature (see the discussion in LT).  Of the 9
remaining systems, 5 clearly show soft excesses (Mkn 279 - see
e.g. Weaver, Gelbord \& Yaqoob 2001; NGC 7469 - see e.g. DeRosa,
Fabian \& Piro 2002; PG 0804+761 \& PG 1211+143 - see e.g. George et
al. 2000; Mkn 335 - see e.g. Turner \& Pounds 1988), two show spectra
softer than $\alpha=-0.95$ (PG 0052+251 - LT \& PG 0953+415 - George
et al. 2000), one shows a fairly typical X-ray spectrum (PG 1307+085 -
LT), and only one shows a spectrum harder than the typical quasar
spectrum (PG 1613+658 - LT).

The sources with relatively soft spectra, but no clear soft excess
have higher black hole masses than the sources with clear soft
excesses (pushing their disc component's peak to lower energies since
$T_{disc}\propto$$M_{BH}^{-1/4}$), and are at higher redshifts, so the
observed soft X-rays probe a slightly higher energy in the rest frame.
One thus might expect to need to use the EUV to find their ``soft
X-ray excesses.''  It is worth noting that the single truly hard X-ray
spectrum belongs to a source which also shows the only flat radio
spectrum among the sources (Falcke, Malkan \& Biermann 1995; Ho 2002)
and has an inferred luminosity just barely higher than 2\% of
$L_{EDD}$; it may represent a source placed into the wrong bin in the
correlation due to measurement errors or hysteresis.  Thus while we
have applied ``mix-and-match'' criteria to discussion the spectra,
there seems to be fairly suggestive anecdotal evidence that the
spectra of the systems which are well below the GFP03, MHD03 \& FKM03
correlation curves and which lie in the 2-10\% of $L_{EDD}$ range have
systematically softer X-ray spectra than the sources; the correlations
between X-ray hardness and radio loudness found by Elvis et al. (1994)
and by Zdziarski et al. (1995) using the classical definition of radio
loudness (i.e. the radio to optical flux ratio) rather than the
radio-to-bolometric luminosity ratio hold up.

\section{Acknowledgments}
TM thanks Mike Eracleous for useful discussions which helped to
stimulate this work.  We thank the referee, Andrea Merloni, for useful
suggestions which improved this work.

\label{lastpage}
\end{document}